\documentclass[aps,pre,twocolumn,groupedaddress]{revtex4}

\usepackage{graphicx}

\begin{document}

\title{Simulations of Helix Unwinding in Ferroelectric Liquid Crystals}

\author{Nurit Baytch}
\altaffiliation{Current address:  Physics Department, Massachusetts Institute of
Technology, Cambridge, MA 02139.}
\author{Robin L. B. Selinger}
\affiliation{Physics Department, Catholic University of America,
Washington, DC 20064}
\author{Jonathan V. Selinger}
\author{R. Shashidhar}
\altaffiliation{Current address:  Geo-Centers Inc., Maritime Plaza One, Suite
050, 1201 M Street, SE, Washington, DC 20003.}
\affiliation{Center for Bio/Molecular Science and Engineering,
Naval Research Laboratory, Code 6900,\\
4555 Overlook Avenue, SW, Washington, DC 20375}

\date{July 28, 2003}

\begin{abstract}
In bulk ferroelectric liquid crystals, the molecular director twists in a helix.
In narrow cells, this helix can be unwound by an applied electric field or by
boundary effects.  To describe helix unwinding as a function of both electric
field and boundary effects, we develop a mesoscale simulation model based on a
continuum free energy discretized on a two-dimensional lattice.  In these
simulations, we determine both the director profile across the cell and the net
electrostatic polarization.  By varying the cell size, we show how boundary
effects shift the critical field for helix unwinding and lower the saturation
polarization. Our results are consistent with experimental data.
\end{abstract}


\maketitle

\section{Introduction}

One of the most extensively studied phases of liquid crystals, both for basic
research and for applications, is the smectic-C* (SmC*) phase of chiral
molecules.  In this phase, the molecules lie in layers and are tilted with
respect to the layer normal direction.  The combination of molecular chirality,
smectic layering, and tilt order leads to two effects:  a ferroelectric
polarization within the smectic layer plane and a helical modulation in the
orientation of the molecular tilt from layer to layer~\cite{degennes93}.  The
ferroelectric polarization is useful for display devices, which use an applied
electric field to switch the molecular orientation~\cite{goodby91}.  It is also
useful for thermal sensors, which measure the temperature variation of the
polarization, known as the pyroelectric effect~\cite{osullivan95,crandall99}.
Both of these applications require a uniform orientation of the molecules.
Hence, the helix must be suppressed, or unwound, by an applied electric field or
by boundary effects.

Helix unwinding has been modeled through continuum elastic theory.  In a bulk
SmC* phase, unwinding induced by an applied electric field can be described by
the sine-Gordon equation, presented below~\cite{meyer68,meyer69,kamien01}.
Under an electric field, the helix distorts and the helical pitch increases.  If
the field is increased above a critical threshold, the pitch diverges and the
helix is suppressed.  By contrast, helix unwinding induced by boundary effects
in a narrow cell is more complex~\cite{cladis72,luban74,glogarova83,povse93}.
In this case, a helix must have a series of disclination lines near the cell
surfaces, which separate a twisted interior region from a uniform surface
region.  Continuum elastic theory shows that the unwinding transition is
controlled by an energetic competition between the helical state with
disclinations and the uniform untwisted state.  If the cell thickness is below a
critical threshold comparable to the helical pitch, the uniform state is favored
and the helix is suppressed.  (A third possible mechanism for helix unwinding is
shear flow~\cite{rey96}, but this is not often used in SmC* liquid crystals, and
we will not discuss it here.)

In this paper, we investigate the unwinding of a SmC* helix by boundary effects,
or by combined boundary and electric-field effects, through a series of Monte
Carlo simulations of a continuum free energy discretized on a 2D lattice.  These
simulations serve two purposes.  First, they allow us to visualize the complex
director configuration within a narrow cell as a function of cell thickness and
electric field.  We obtain snapshots of the molecular tilt profile through the
helix unwinding process.  Second, they allow us to relate the microscopic
director configuration to two macroscopic variables, an average chiral order
parameter and the net electrostatic polarization of the cell.  The latter
variable can be compared with experimental measurements of SmC* cells.

In these simulations, we first consider boundary effects alone.  We simulate the
SmC* phase in narrow cells, and determine the structure of the helix as a
function of cell thickness.  We confirm that the helix unwinds at a threshold
thickness approximately equal to the helical pitch, in agreement with continuum
elastic theory.  We then use the same Monte Carlo approach to simulate helix
unwinding due to the combined effects of cell boundaries and electric field.  In
these simulations, we calculate both the molecular tilt profile across the cell
and the net electrostatic polarization.  We observe three distinct regimes of
response, in which the helical pitch is first distorted and then expelled as the
applied field is increased.  By varying the cell thickness, we determine how
boundary effects shift the critical field for helix unwinding and lower the
saturation polarization.  These simulation results are consistent with trends
observed in experiments.

The plan of this paper is as follows.  In Sec.~II, we review the continuum
elastic theory of the SmC* phase.  We show how an applied electric field unwinds
the SmC* helix in a bulk system, and sketch an approximate energetic argument
for a finite cell.  In Sec.~III, we use Monte Carlo simulations to investigate
the effects of finite cell thickness under zero applied electric field.  In
Sec.~IV, we combine the effects of cell boundaries and applied electric field to
determine the tilt profile and the polarization, and compare the polarization
with experimental measurements.

\section{Theory}

\begin{figure}
\includegraphics[clip,height=3in]{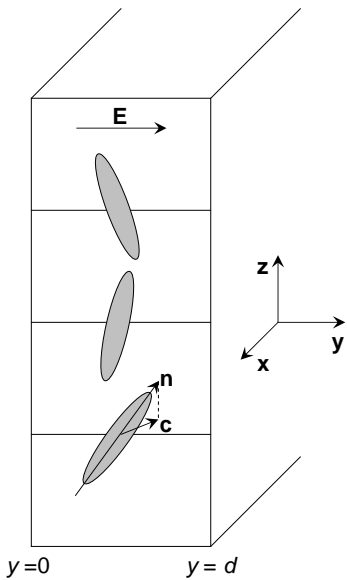}%
\caption{Idealized bookshelf geometry of a SmC* liquid crystal
in a narrow cell.  The helical axis is along the $z$-axis, perpendicular to the
smectic layers.  An electric field is applied along the $y$-axis.}
\label{geometry}
\end{figure}

A SmC* liquid crystal in a narrow cell has the idealized bookshelf geometry
shown in Fig.~\ref{geometry}.  The molecules lie in layers, and they are tilted
with respect to the smectic layer normal.  The 3D orientation is represented by
the director $\mathbf{n}$.  This unit vector can be written in spherical
coordinates as $\mathbf{n}=(\sin\theta\cos\phi,\sin\theta\sin\phi,\cos\theta)$,
where $\theta$ is the polar angle of the tilt and $\phi$ is the azimuthal angle.
The molecular tilt is conventionally described in terms of the projection of
$\mathbf{n}$ into the layer plane,
$\mathbf{c}=(\sin\theta\cos\phi,\sin\theta\sin\phi)$.

In this geometry, we expect the director to depend on the $y$ and $z$
coordinates.  The $z$ coordinate is along the smectic layer normal.  Because of
the molecular chirality, the molecular orientation rotates in a helix from layer
to layer, which makes the director a periodic function of $z$.  The $y$
coordinate is the narrow dimension of the cell, with a thickness $d$ of order
microns, across which an electric field $\mathbf{E}$ is applied.  The molecules
may interact strongly with the front and back surfaces of the cell, at $y=0$ and
$d$.  As a result, the director may rotate as a function of $y$.  By contrast,
we do not expect the director to depend on the third coordinate $x$.  In this
bookshelf geometry, the system is uniform as a function of $x$. (In certain
liquid crystals, the smectic layers buckle as a function of
$x$~\cite{crawford94,geer98,bartoli98,selinger00}, but we do not consider that
effect here.)  For that reason, we write the angles $\theta$ and $\phi$ as
functions of $y$ and $z$.

We can now construct the simplest model free energy to describe variations in
the director.  This model must include four interactions.  First, the smectic
layer order interacts with the molecular orientation, and favors a particular
tilt of the molecules with respect to the layers.  This interaction can be
expanded as a power series in the tilt magnitude $|\mathbf{c}|=\sin\theta$,
which gives $-\frac{1}{2}r|\mathbf{c}|^2+\frac{1}{4}u|\mathbf{c}|^4$, for some
series coefficients $r$ and $u$.  In terms of these coefficients, the favored
tilt is $|\mathbf{c}|=(r/u)^{1/2}$.  Second, there is an electroclinic
interaction of the molecules with the applied electric field $\mathbf{E}$.
Because of the molecular chirality, this interaction couples the field in the
$y$ direction with the tilt in the $x$ direction, so it can be written as
$b\mathbf{z}\cdot\mathbf{E}\times\mathbf{c}=-b E_y c_x$.  In other words, the
electrostatic polarization is $\mathbf{P}=b\mathbf{z}\times\mathbf{c}$, or
$P_y=-b c_x$. Third, there is a chiral interaction that favors a variation of
the director from layer to layer, which can be written as
$-\lambda\mathbf{z}\cdot\mathbf{c}\times\partial\mathbf{c}/\partial z$.
Fourth, there is the Frank free energy for elastic distortions of the director,
which limits the variations from layer to layer.  This contribution can be
written as $\frac{1}{2}K(\partial_i c_j)(\partial_i c_j)$, summed over $i$ and
$j$.  Putting these pieces together, the total free energy density becomes
\begin{equation}
F=-\frac{1}{2}r|\mathbf{c}|^2+\frac{1}{4}u|\mathbf{c}|^4
+b\mathbf{z}\cdot\mathbf{E}\times\mathbf{c}
-\lambda\mathbf{z}\cdot\mathbf{c}\times\frac{\partial\mathbf{c}}{\partial z}
+\frac{1}{2}K(\partial_i c_j)(\partial_i c_j).
\label{freeenergy}
\end{equation}
This free energy is invariant under rotations in the $xy$ plane, but it is not
invariant under reflections, because reflection symmetry is broken by the
molecular chirality.

\begin{figure}
\includegraphics[clip,width=3.375in]{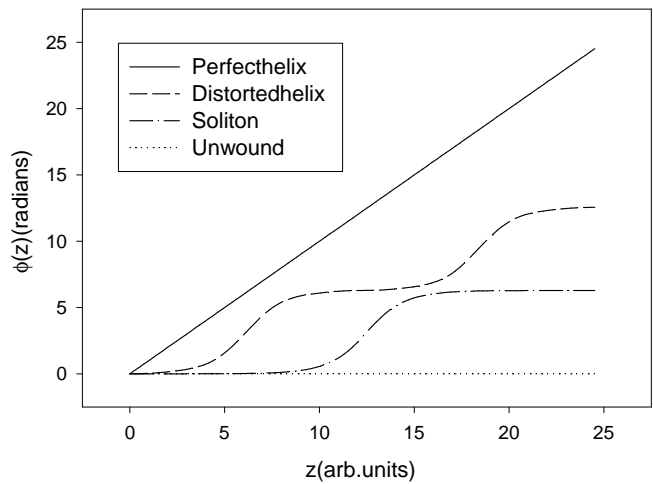}%
\caption{Unwinding of a SmC helix in a bulk system under an applied electric
field $E$.  Solid line:  Perfect helix for $E=0$.  Dashed line:  Distorted helix
under moderate $E$.  Dot-dashed line:  Single soliton in the director, near the
threshold for helix unwinding.  Dotted line:  Unwound uniform director.}
\label{helix}
\end{figure}

We can see immediately that this minimal model leads to a helix in the $z$
direction.  Consider a director of the form $\mathbf{c}=(c\cos\phi,c\sin\phi)$,
where the magnitude $c=\sin\theta$ is constant and the azimuthal angle $\phi$
depends on position.  The free energy density simplifies to
\begin{equation}
F=-\frac{1}{2}r c^2 +\frac{1}{4}u c^4-b E c \cos\phi
-\lambda c^2 \frac{\partial\phi}{\partial z}
+\frac{1}{2}K c^2 \left|\nabla\phi\right|^2.
\end{equation}
In the limit of zero electric field, $E=0$, we minimize this free energy over
$\phi$ to obtain
\begin{equation}
\frac{\partial\phi}{\partial z}=\frac{\lambda}{K}.
\end{equation}
Hence, the azimuthal angle increases linearly as $\phi(z)=\phi_0+q_0 z$, with
$q_0=\lambda/K$, as shown by the solid line in Fig.~\ref{helix}.  This linear
increase in $\phi$ implies a perfect sinusoidal helix in $\mathbf{c}$.

This minimal model also predicts unwinding of the helix under an applied
electric field.  This behavior is analogous to the theory of Meyer for a
cholesteric phase in an electric or magnetic
field~\cite{meyer68,meyer69,kamien01}.  For $E\not=0$, minimizing the free
energy over $\phi$ gives
\begin{equation}
\frac{\partial^2\phi}{\partial z^2}=\frac{b E}{K c}\sin\phi.
\end{equation}
This is the standard sine-Gordon equation.  The form of the solutions depends on
the value of $E$, as shown in Fig.~\ref{helix}.  For low $E$, the helix is
distorted, so that the director is approximately aligned with the field in most
of the system.  As $E$ increases, the helix becomes even more distorted, with
sharper steps between domains where $\phi$ is approximately a multiple of
$2\pi$.  Eventually the system crosses over into regime of uniform domains
separated by sharp domain walls, or solitons.  In that regime, a single soliton
has the profile $\phi(z)=2\pi-4\tan^{-1}[\exp(-(z-z_{\rm wall})/\xi)]$, where
$\xi=(2 K c / b E)^{1/2}$.  As $E$ continues to increase, the spacing between
the domain walls increases, or equivalently their density decreases.  At the
critical threshold $E=(\pi^2/8)(\lambda^2 c/b K)$, the last domain wall vanishes
and the system becomes uniform.

The question is now:  What other types of influences can also unwind a SmC*
helix?  Clearly one possibility is surface effects.  Interactions along the
front and back surfaces of a finite cell, at $y=0$ and $d$, can anchor the
director at those surfaces.  If the elastic interactions described by the
parameter $K$ are sufficiently strong, and the cell thickness $d$ is not too
big, then this anchoring may extend throughout the interior of the cell, giving
a uniform director.  This is the basis of surface-stabilized ferroelectric
liquid crystal cells~\cite{goodby91}.

The threshold thickness for unwinding a helix is not obvious.  As shown in
Fig.~\ref{geometry}, the helical pitch is along the $z$ direction, but the
narrow dimension of the cell is along the $y$ direction.  Because these
directions are perpendicular, there is no simple geometric reason why a helix
must unwind when the cell thickness is less than the pitch.  Rather, there must
be some energetic argument that relates these two length scales.

An energetic argument has been developed for narrow cells of the cholesteric
phase~\cite{cladis72,luban74}, and has been extended to narrow cells of the SmC*
phase~\cite{glogarova83,povse93}.  In its simplest form, the argument can be
stated as  follows~\cite{meyerprivate}.  If a cell has a helix in the interior,
but a uniform director along the front and back surfaces, then it must have a
series of disclination lines running along the $x$ direction near the surfaces.
There must be one disclination line for each helical pitch.  We can compare the
energy of the helix (with disclinations) with the energy of the uniform state to
find the threshold thickness for helix unwinding.  The energy of the helix (with
disclinations) is the negative energy gained from the helix plus the positive
energy lost to the disclinations,
\begin{equation}
\frac{\Delta E}{\text{volume}} \approx -K q_0^2 +
\frac{E_{\text{line}}}{(d)(\text{pitch})},
\end{equation}
where $E_{\text{line}}$ is the disclination line energy per unit length.  The
helix unwinds if $\Delta E >0$, which implies
\begin{equation}
d\gtrsim\left(\frac{E_{\text{line}}}{K}\right)(\text{pitch}).
\label{thicknessthreshold}
\end{equation}
Since the line energy $E_{\text{line}}$ should be of order $K$, the threshold
thickness should be comparable to the pitch.

In the following sections, we test this argument through a series of Monte Carlo
simulations.  In these simulations, we obtain snapshots of the director
configuration for different cell thicknesses, both at zero field and under an
applied electric field.  These snapshots provide specific illustrations of the
disclinations discussed above.
For zero field, the simulations confirm that the helix
unwinds at a critical thickness approximately equal to the helical pitch.  For
finite field, the simulations show helix unwinding induced by the combined
effects of boundaries and electric field in a cell above the critical thickness.
In both cases, we relate the microscopic snapshots of the director configuration
to macroscopic variables.  One of these variables, the net electrostatic
polarization, can be compared with experimental measurements.

\section{Finite cells under zero electric field}

To model helix unwinding in a finite cell of the SmC* phase, we map the system
onto a 2D square lattice.  The lattice dimensions represent the $yz$ plane shown
in Fig.~\ref{geometry}:  $y$ is the narrow dimension of the cell and $z$ is the
smectic layer normal.  We assume the system is uniform in the $x$ direction.  On
each lattice site $(y,z)$ there is a 3D unit vector ${\mathbf n}(y,z)$
representing the local molecular director.  This vector can be parametrized in
terms of the polar angle $\theta(y,z)$ and azimuthal angle $\phi(y,z)$, or
equivalently in terms of the projection ${\mathbf c}(y,z)$ into the smectic
layer plane.

For the lattice simulations, we discretize the free energy of
Eq.~(\ref{freeenergy}) to obtain
\begin{eqnarray}
F&&=\sum_{(y,z)}\biggl[-\frac{1}{2}r|\mathbf{c}(y,z)|^2
+\frac{1}{4}u|\mathbf{c}(y,z)|^4
+b\mathbf{z}\cdot\mathbf{E}\times\mathbf{c}(y,z)\nonumber\\
&&
-\lambda\mathbf{z}\cdot
\frac{\mathbf{c}(y,z)+\mathbf{c}(y,z+1)}{2}
\times
[\mathbf{c}(y,z+1)-\mathbf{c}(y,z)]\\
&&
+\frac{1}{2}K(|\mathbf{c}(y+1,z)-\mathbf{c}(y,z)|^2
+|\mathbf{c}(y,z+1)-\mathbf{c}(y,z)|^2)
\biggr] ,\nonumber
\label{latticefreeenergy}
\end{eqnarray}
with $\mathbf{E}=0$ in this section.  This free energy is similar to the free
energy studied in Ref.~\cite{selinger00} but with one important difference:
that paper simulated the $xy$ plane, but we now simulate the $yz$ plane.

A further consideration for the simulations is boundary conditions.
Experimental cells may be symmetric, with the local director aligned along the
same direction on both front and back confining walls, or they may be
asymmetric, with the director aligned along one wall and an open boundary on the
other side.  In our simulations we use an aligning boundary condition with the
director fixed on the wall at $y=0$ with a specified tilt angle.  On the other
wall at $y=d$ we use the boundary condition  $\partial c(y,z)/\partial y=0$.
This arrangement can represent an asymmetric cell, or one half of a symmetric
cell, with the other half a mirror image of the first.  Experimental cells are
very large in the $z$ direction so that the top and bottom boundaries should not
affect the physics of the interior.  In the simulations, we use the boundary
condition $\partial c(y,z)/\partial z=0$ for the top and bottom boundaries.

We simulate the system with the parameters $r=0.007596$, $u=1$, $b=1$, $E=0$,
$K=1.5$, and $\lambda=0.25$ and $0.125$.  The small value of $r$ corresponds to
a tilt angle of approximately $5^\circ$.  We use the large $z$
dimension of 160, and several values of the thickness $d$ in the $y$ direction.
For each set of parameters and system size, we begin the simulations in a
disordered state at a high Monte Carlo temperature, and then slowly reduce the
temperature until the system settles into a single ground state and the
fluctuations in $\mathbf{c}$ become negligible.  This procedure can be regarded
as a simulated-annealing minimization of the lattice free energy of
Eq.~(\ref{latticefreeenergy}).

\begin{figure}
(a)\includegraphics[clip,height=2.27in]{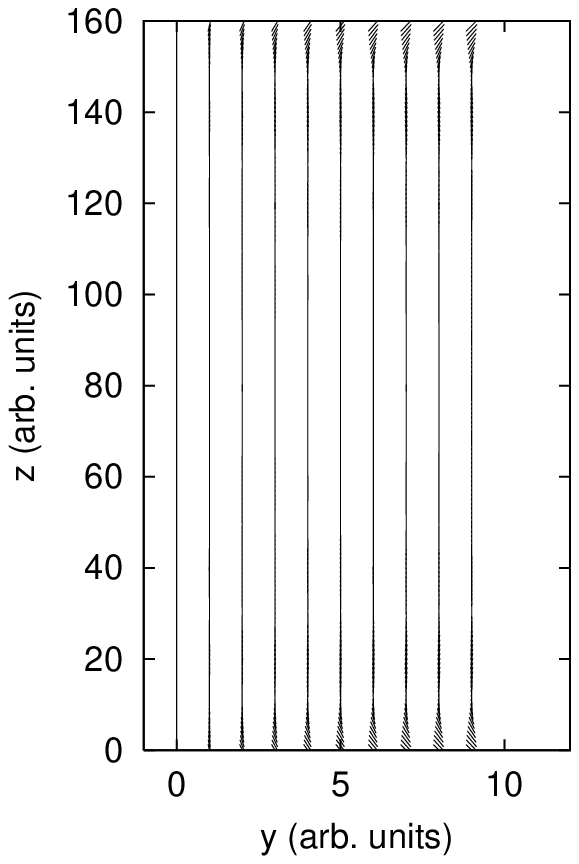}\ %
(b)\includegraphics[clip,height=2.27in]{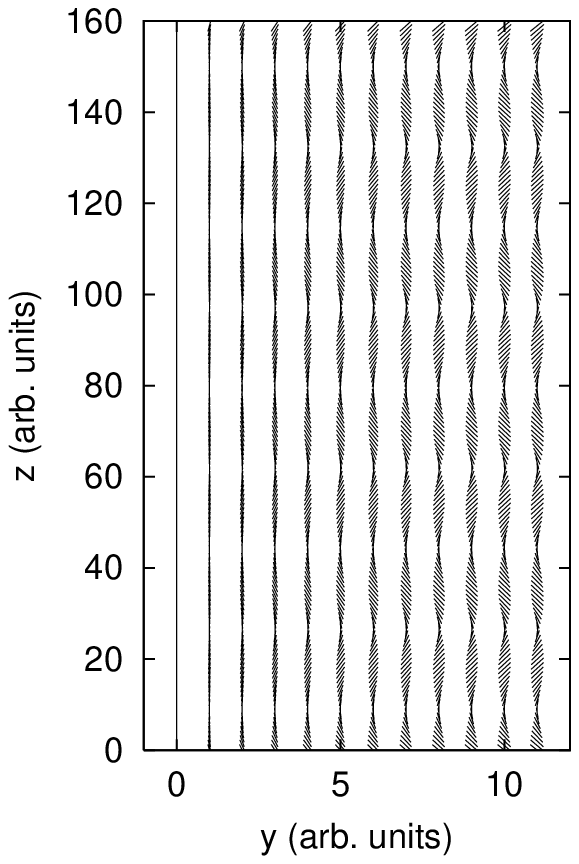}%
\caption{Director configuration in the $yz$ plane for simulations with
$\lambda=0.25$, leading to a pitch of approximately 38.  The system has an
aligning boundary along the left side ($y=0$).  Note that the $y$ axis is
exaggerated compared with the $z$ axis.  (a)~For a thickness of 10 ($y=0$
through $9$), the system is uniform, except for some edge effects at the top
and bottom.  (b)~For a thickness of 12, the system has a clear helix.}
\label{lambda25}
\end{figure}

\begin{figure}
(a)\includegraphics[clip,height=2.27in]{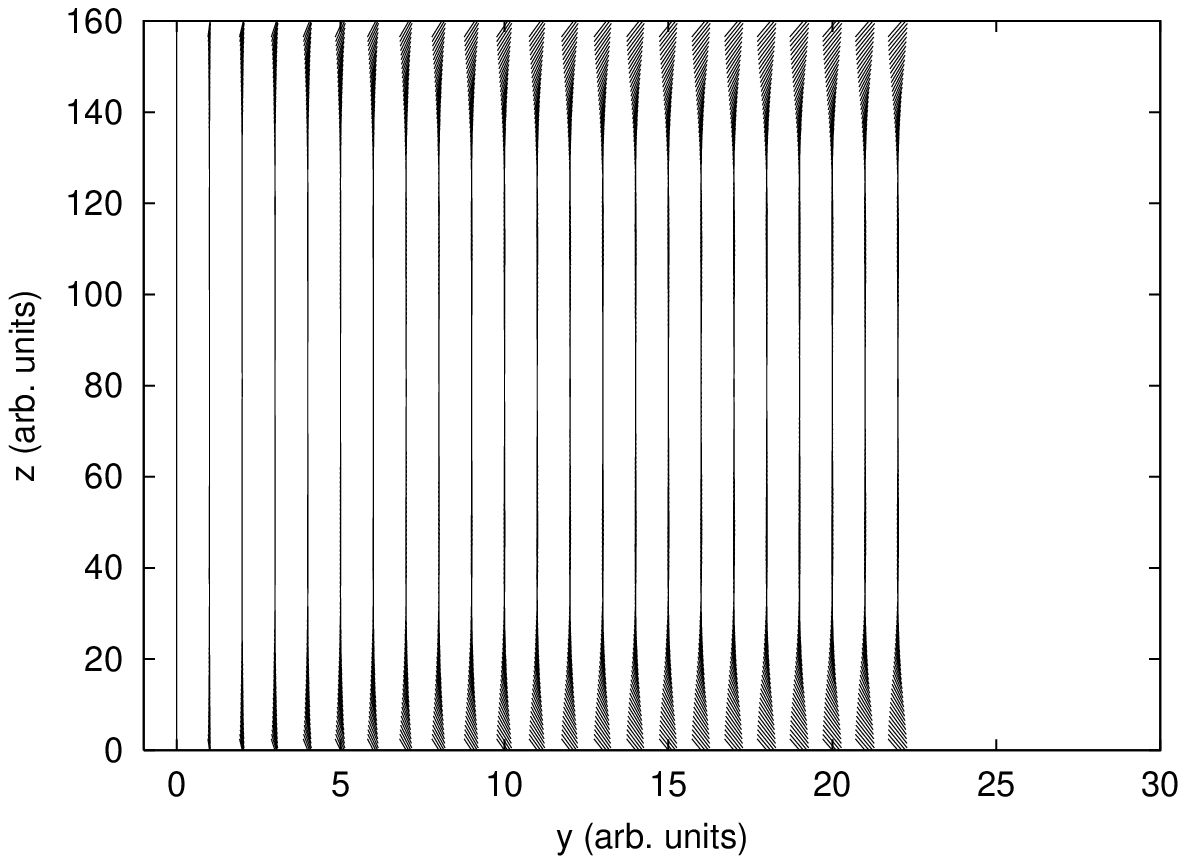}%

(b)\includegraphics[clip,height=2.27in]{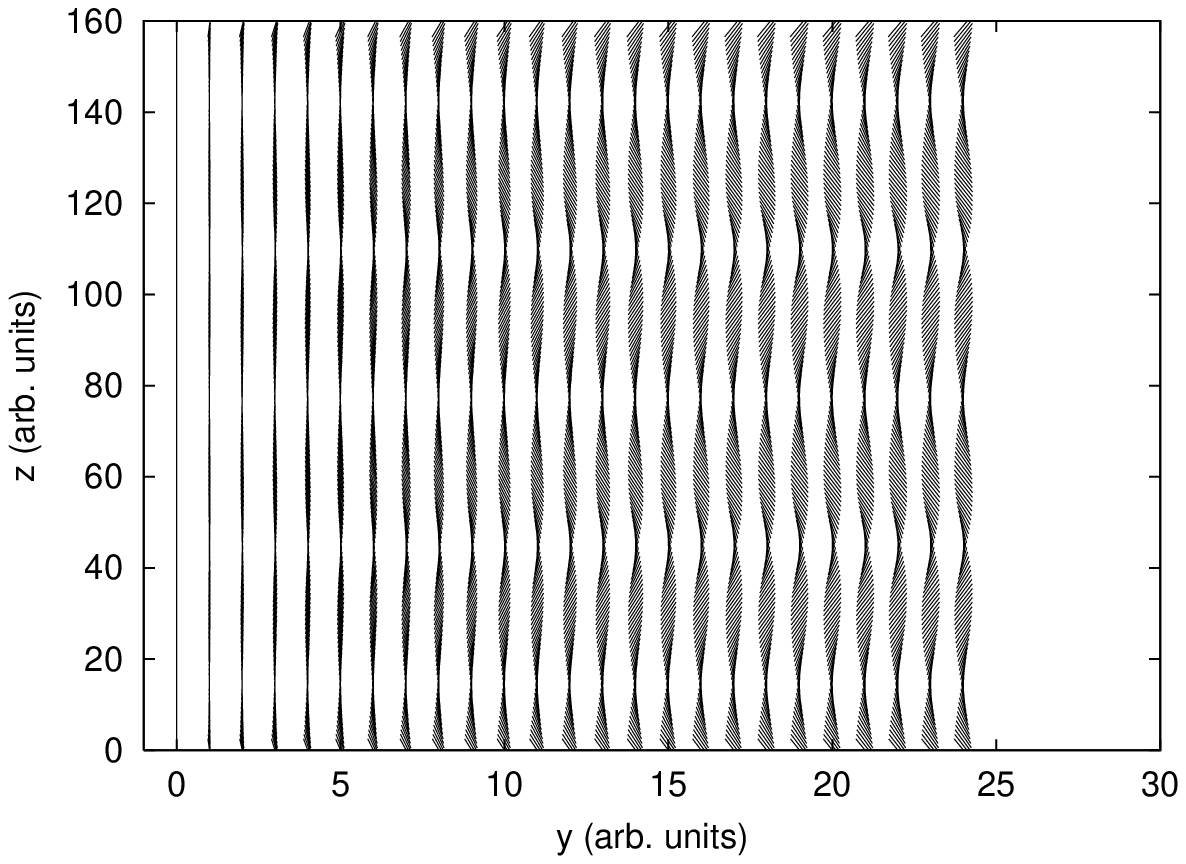}%
\caption{Director configuration in the $yz$ plane for simulations with
$\lambda=0.125$, leading to a pitch of approximately 75.  (a)~For a thickness of
23, the system is uniform, except for some edge effects.  (b)~For a thickness of
25, there is a clear helix.}
\label{lambda125}
\end{figure}

To visualize the simulation results, we draw the $yz$ plane in Figs.
\ref{lambda25} and \ref{lambda125}.  The director configuration is represented
by short lines that show the projection of the 3D director into the $yz$ plane.
Hence, vertical lines indicate $c_y=0$, and tilted lines indicate $c_y\not=0$.
Because the lines representing the director are drawn close together, helical
regions resemble twisted ribbons.

Figure~\ref{lambda25} shows the simulation results for $\lambda=0.25$ and
$K=1.5$.  For these parameters, the favored wavevector is
$q_0=\lambda/K\approx0.17$, and hence the unperturbed pitch is
$2\pi/q_0\approx38$.  For a thickness of 10, the system is uniform, with $c_y=0$
everywhere except near the top and bottom surfaces.  Those distortions are edge
effects within a fractional pitch of the free surfaces, which do not affect the
bulk behavior inside the cell.  Hence, we see that the system of thickness 10
is unwound.  By contrast, for a thickness of 12, the system shows a well-defined
helix, with a periodic modulation of $c_y$ throughout the cell, except very
close to the aligning surface at $y=0$.  Thus, there is a clear helix
winding/unwinding transition as a function of thickness.  The transition occurs
at a critical thickness between 10 and 12 for a cell with asymmetric boundary
conditions (or a half-thickness between 10 and 12 if the simulation is regarded
as half of a symmetric cell).  This critical thickness is not equal to the
pitch, but it is certainly the same order of magnitude, in agreement with the
theoretical expectation of Eq.~(\ref{thicknessthreshold}).

Figure~\ref{lambda125} shows the corresponding results for $\lambda=0.125$.  In
this case, the favored wavevector is $q_0\approx0.083$, so the unperturbed pitch
is $2\pi/q_0\approx75$.  For a thickness of 23, the system is uniform, again
except for some edge effects near the top and bottom surfaces.  By comparison,
for a thickness of 25, there is a distinct helix throughout the interior of the
cell.  Thus, there is a helix winding/unwinding transition with a
critical thickness between 23 and 25.  This critical thickness is approximately
twice the critical thickness of the previous case.  Hence, we see that the
critical thickness is approximately proportional to the pitch, again in
agreement with the theoretical expectation.

\begin{figure}
\includegraphics[clip,width=3.375in]{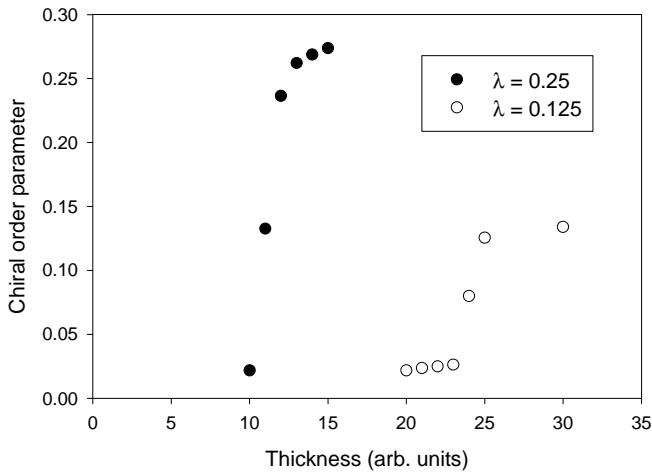}%
\caption{The chiral order parameter $\chi$ defined in
Eq.~(\protect{\ref{orderparameter}}), as a function of the system thickness $d$.
For $\lambda=0.25$ (pitch $\approx 38$), the unwinding transition occurs at a
thickness between 10 and 12.  For $\lambda=0.125$ (pitch $\approx 75$), the
transition occurs at a thickness between 23 and 25.}
\label{orderparameterfig}
\end{figure}

For a quantitative measurement of helix winding and unwinding, we must define a
chiral order parameter.  One simple choice of a chiral order parameter is just
the chiral term of the free energy~(\ref{latticefreeenergy}), without the factor
of $\lambda$ itself,
\begin{eqnarray}
\chi&=&-\frac{1}{N_{\text{sites}}}\sum_{(y,z)}
\mathbf{z}\cdot\frac{\mathbf{c}(y,z)+\mathbf{c}(y,z+1)}{2}\nonumber\\
&&\phantom{-\frac{1}{N_{\text{sites}}}\sum_{(y,z)}}
\times[\mathbf{c}(y,z+1)-\mathbf{c}(y,z)].
\label{orderparameter}
\end{eqnarray}
Figure~\ref{orderparameterfig} shows this order parameter as a function of the
system thickness $d$ for both series of simulations, with $\lambda=0.25$ and
$0.125$.  For $\lambda=0.25$, the plot shows a sharp transition between
thickness 10 and 12, as $\chi$ jumps from 0.022 to 0.24.  For
$\lambda=0.125$, there is a distinct transition between
thickness 23 and 25, as $\chi$ jumps from 0.026 to 0.13.  This analysis
confirms that the winding/unwinding transition occurs at a thickness that is
proportional to, and the same order of magnitude as, the helical pitch.

\section{Finite cells under an electric field}

In the previous section, we showed that a SmC* helix can be unwound by surface
effects in a finite cell.  If the thickness is greater than the critical
threshold, the helix is present at zero electric field.  However, when an
electric field is applied, the helix can be unwound by the combined effects of
the surfaces and the electric field.  In this section, we simulate that
combination of surface and field effects.

For these simulations, we use the same Monte Carlo approach as in the previous
section.  We use the discretized free energy of Eq.~(\ref{latticefreeenergy})
with the parameters $r=0.0625$, $u=1$, $b=1$, $K=1.5$, and $\lambda=0.25$.  This
value of $r$ corresponds to a tilt angle of approximately $15^\circ$.  We
perform the simulations for four values of the cell thickness, $d=10$, 20, 40,
and 60, and scan through the electric field $E$ at each thickness.  For these
parameters, the unperturbed pitch is $2\pi K/\lambda\approx38$ and hence, based
on the results of the previous section, the zero-field winding/unwinding
transition occurs at a thickness between 10 and 12.  Hence, the simulations at
thickness $10$ should be unwound at all values of the electric field, while the
simulations at larger thickness should go through the winding/unwinding
transition as a function of electric field.

\begin{figure*}
(a)\includegraphics[clip,height=2.27in]{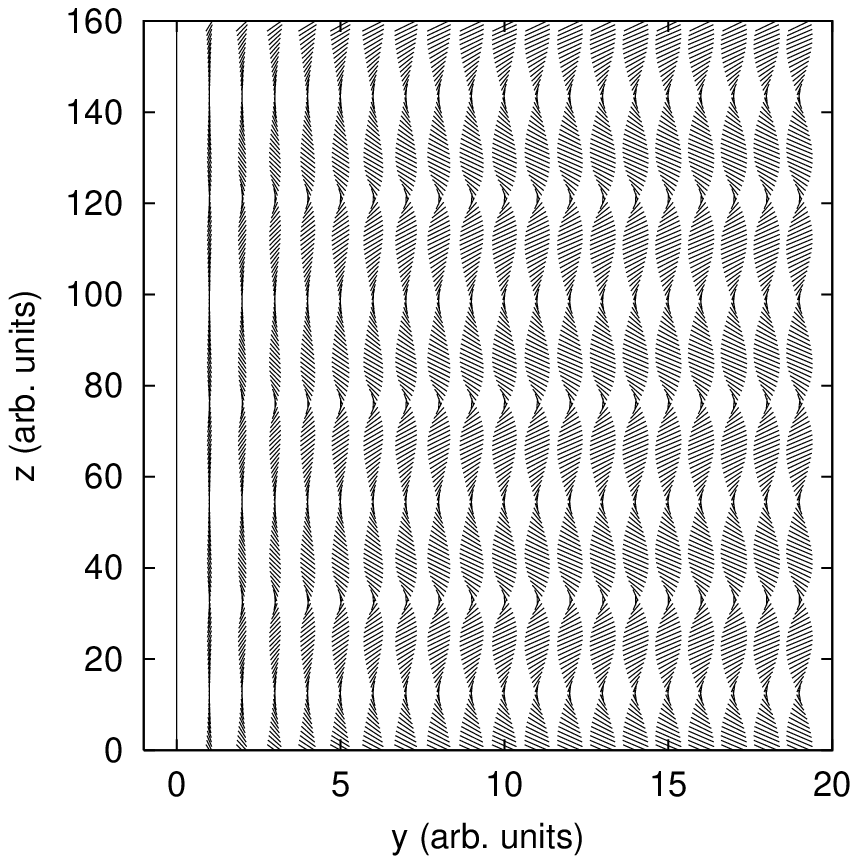}\ %
(b)\includegraphics[clip,height=2.27in]{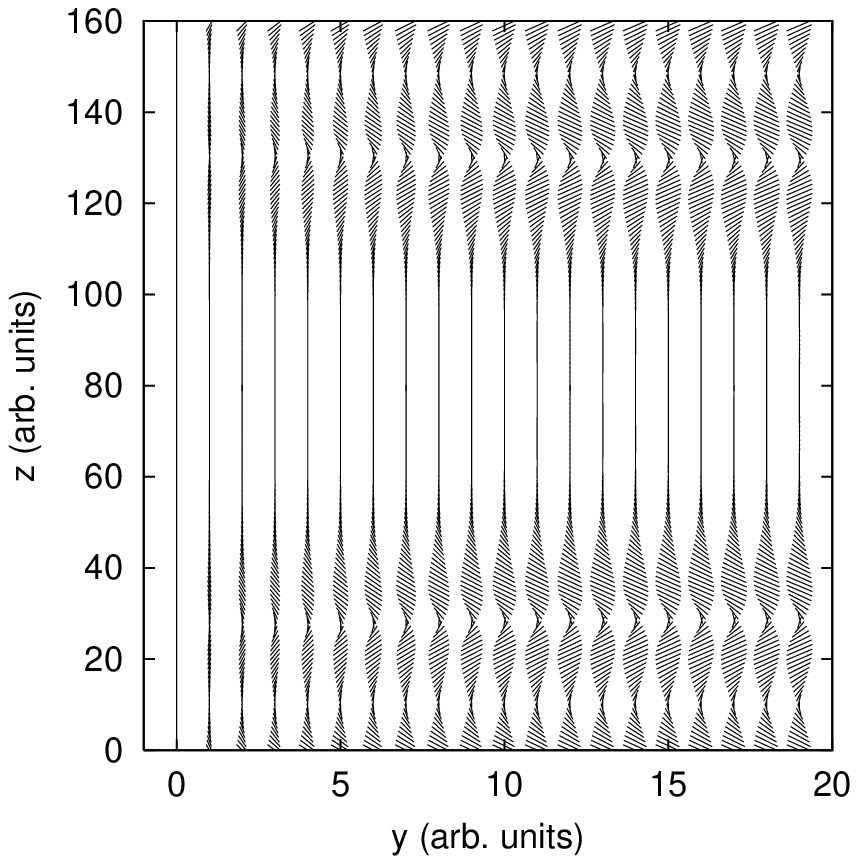}%

(c)\includegraphics[clip,height=2.27in]{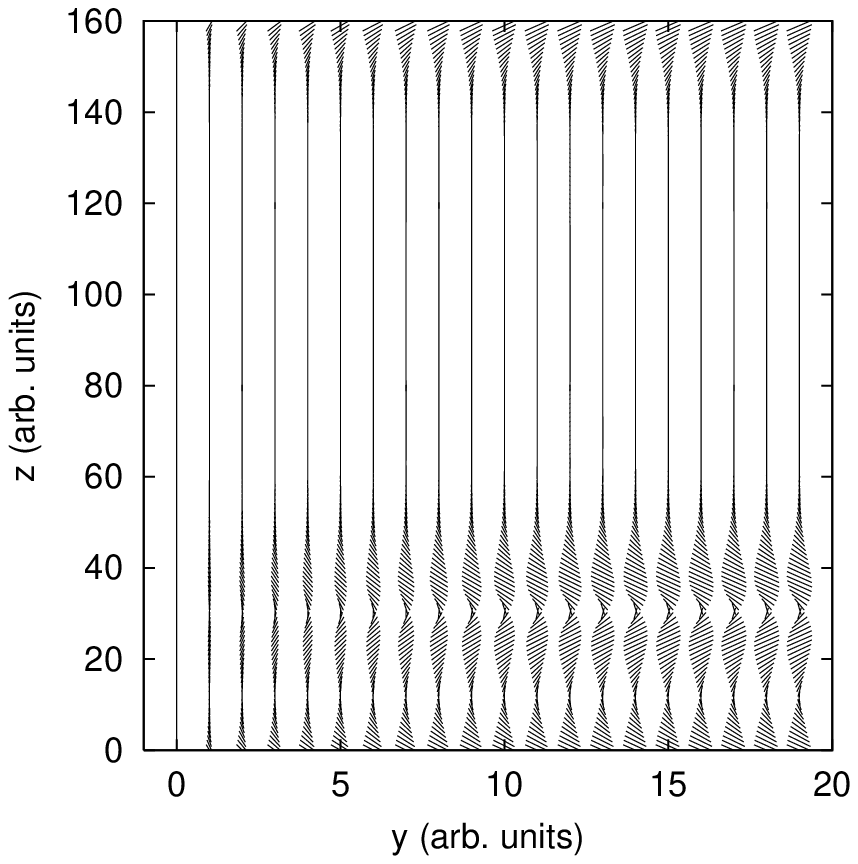}\ %
(d)\includegraphics[clip,height=2.27in]{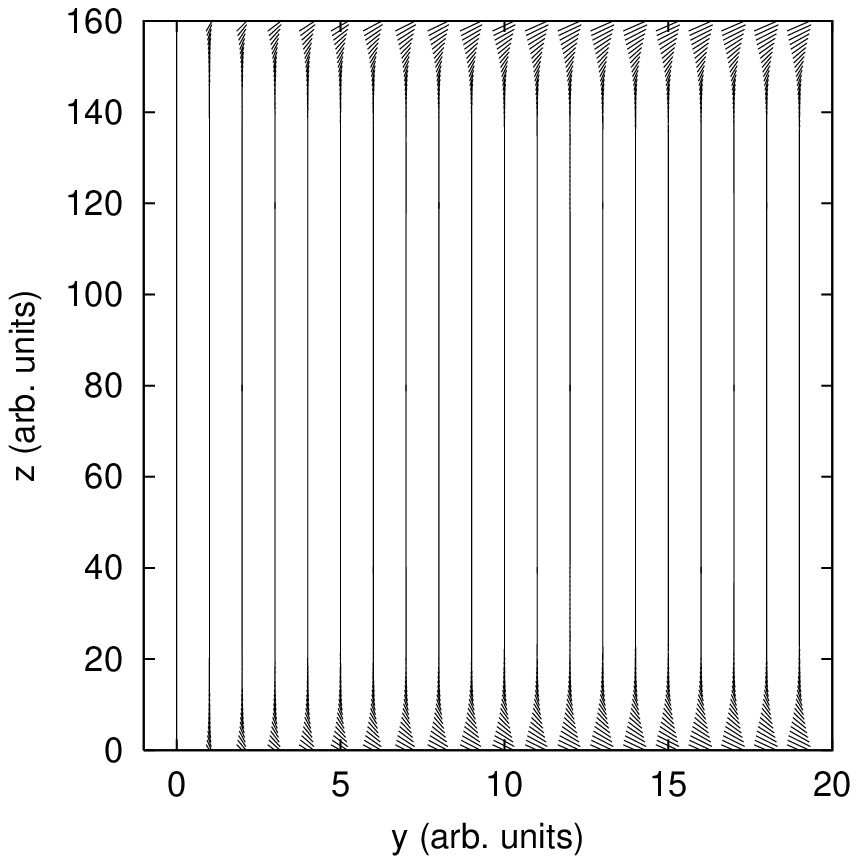}%
\caption{Director configuration in the $yz$ plane as a function of increasing
electric field, for a fixed thickness of 20.  (a)~$E=0$.  (b)~$E=0.004$.
(c)~$E=0.006$.  (d)~$E=0.008$.}
\label{varyfield}
\end{figure*}

Figure~\ref{varyfield} shows the simulation results for the system of thickness
20.  For $E=0$, the system has a helix everywhere in the cell, except a narrow
region near the aligning boundary at $y=0$.  By comparison, for $E=0.004$, the
helix is suppressed in much of the cell.  It persists only in regions of the
cell near the free surfaces at the top and bottom.  This behavior near the free
surfaces is consistent with the previous section, which showed that free
surfaces tend to favor the helical modulation.  When the field increases to
$E=0.006$, the helix is suppressed in more of the cell, and it persists only in
smaller regions near the top and bottom surfaces.  Once the field reaches
$E=0.008$, the helix is suppressed throughout the interior of the cell.  The
director in the cell is now uniform, except for very narrow edge effects at the
top and bottom.  Hence, the electric field has driven the finite cell through
the helix winding/unwinding transition.

\begin{figure}
(a)\includegraphics[clip,width=3.2in]{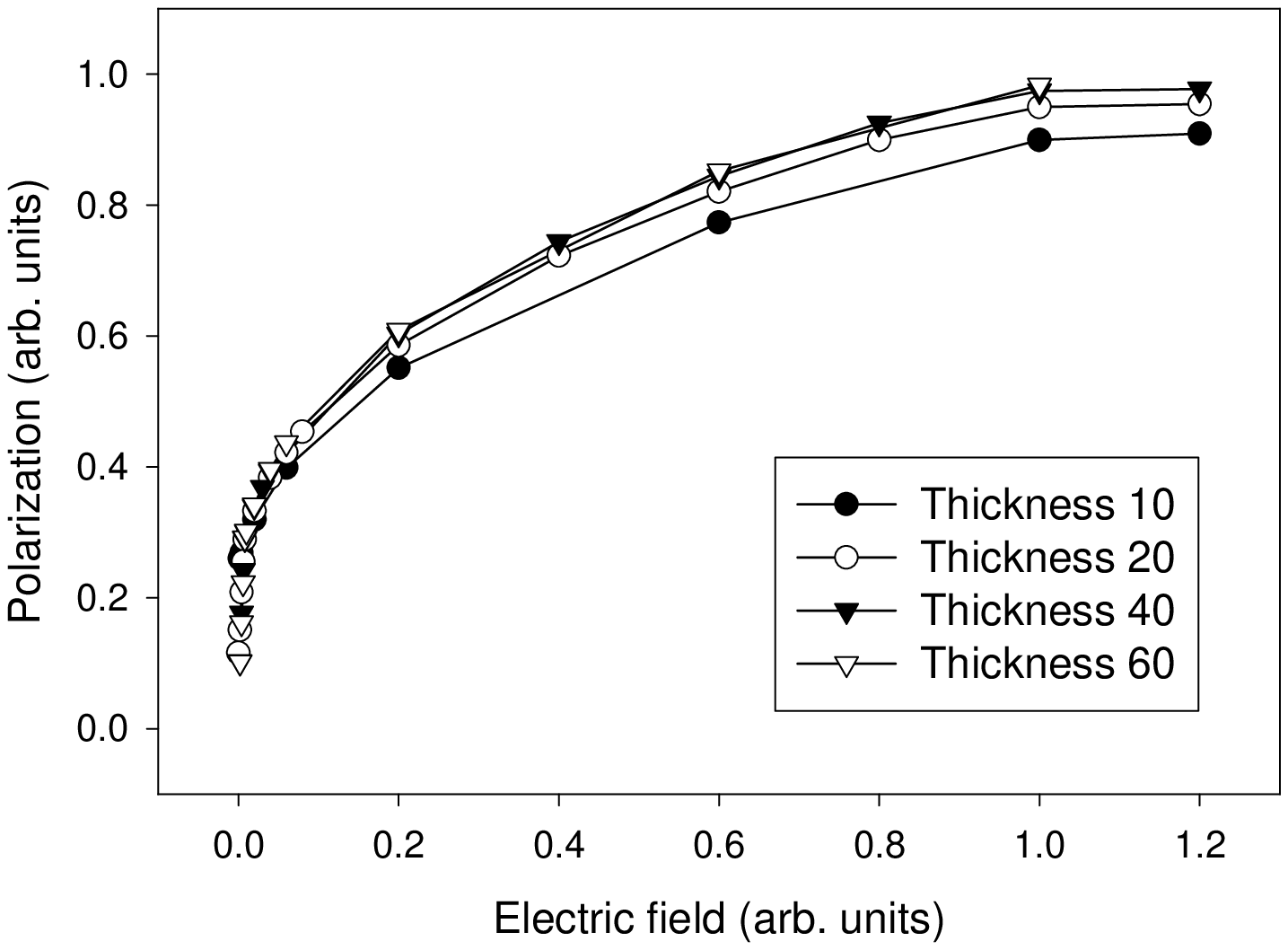}%

(b)\includegraphics[clip,width=3.2in]{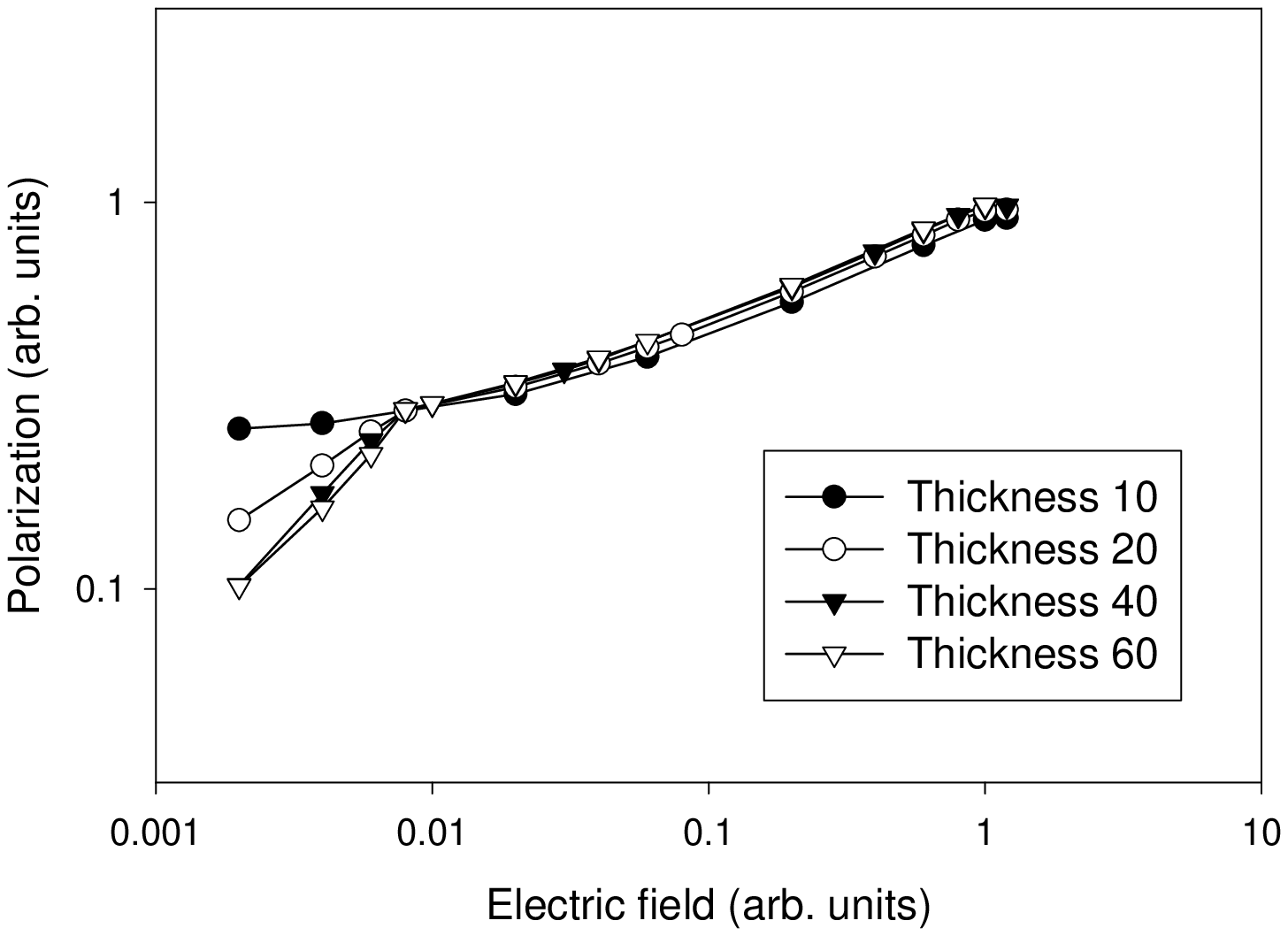}%
\caption{Simulation results for the polarization as a function of electric field
in cells of four thicknesses.  (a)~Linear scale.  (b)~Logarithmic scale.  On the
logarithmic scale, note the three regimes of helix unwinding, suppressed twist,
and saturation.}
\label{pvse}
\end{figure}

As in the previous section, we need an order parameter to describe the
winding/unwinding transition quantitatively.  In this case, the electrostatic
polarization provides an experimentally relevant order parameter, which shows
how the net polar order of the cell couples to the electric field.  As argued in
Sec.~II, the polarization is the quantity conjugate to the electric field in the
free energy, and hence $\mathbf{P}=b\mathbf{z}\times\mathbf{c}$, or $P_y=-bc_x$.
We average this quantity over the cell to obtain
\begin{equation}
P_y=-\frac{1}{N_{\text{sites}}}\sum_{(y,z)}bc_x(y,z).
\label{polarization}
\end{equation}
Figure~\ref{pvse} shows the simulation results for the polarization as a
function of electric field for each cell thickness.  Figure~\ref{pvse}(a)
presents the results on a linear scale, and Fig.~\ref{pvse}(b) presents the same
results on a logarithmic scale.	 Note that the polarization is not zero at zero
field because of the symmetry-breaking surface alignment at $y=0$.

From the plots in Fig.~\ref{pvse}, we can see that the system has three distinct
regimes in its response to an electric field.  For low field $E\lesssim0.01$,
there is a regime of helix unwinding.  In this regime, an increasing electric
field gradually aligns the directors, suppresses the helix, and prevents the
local polarization from averaging to zero.  As a result, the net polarization
increases rapidly as a function of field.  (This low-field regime does not occur
for the lowest thickness $d=10$, because the helix is suppressed by surface
effects even without a field.)  For intermediate field
$0.01\lesssim E\lesssim1$, there is a regime of suppressed twist.  In this
regime, the helix is already unwound, so the only effect of electric field is to
increase the local tilt and polarization.  Hence, the net polarization increases
more slowly, roughly as $E^{1/3}$.  Finally, for high field $E\gtrsim1$, there
is a saturated regime.  Here, the helix is already unwound, the local tilt is at
its maximum value $c_x=1$, and the polarization is at its maximum value of
$P_y=b$.  Although the simulations only go to $E=1.2$, we see that the
polarization cannot increase at higher field because it is already saturated.

In Fig.~\ref{pvse}, we can also compare the relative polarization of thinner and
thicker cells.  For low field, thinner cells have a higher polarization than
thicker cells, because the helix reduces the polarization in thicker cells but
the aligning boundary at $y=0$ unwinds the helix in thinner cells.  By
contrast, for high field, thicker cells have a higher polarization than thinner
cells, because there is no helix at any thickness, and surface effects at $y=d$
suppress the polarization in thinner cells.  A similar high-field limit has been
discussed by Shenoy \emph{et al.}~\cite{shenoy01}, who see the same effect
experimentally in cells with different boundary conditions.  For intermediate
field, the polarization of thinner and thicker cells must cross.

To compare our results with experimental measurements of the polarization as a
function of electric field, we must take into account one subtlety of the
experiments.  As shown by Ruth \emph{et al.}~\cite{ruth94,hermann96}, the
polarization observed experimentally (by the triangle-wave technique or other
techniques) is not the total polarization conjugate to the electric field.
Rather, it is a specific nonlinear component of the polarization, which can be
written as
\begin{equation}
P_{\text{obs}}(E)=P(E)-E\left(\frac{dP}{dE}\right).
\end{equation}
The difference between the total polarization $P(E)$ and the observable
polarization $P_{\text{obs}}(E)$ is small when $P(E)$ is saturated, but it is
significant whenever $P(E)$ varies with $E$.  Hence, we must extract
$P_{\text{obs}}(E)$ from the simulations and compare that quantity with
experiments.

To extract $P_{\text{obs}}(E)$ from the simulations, we need to calculate the
derivative $dP/dE$.  For that reason, we fit the simulation results for the
polarization to the function $P(E)=(1+\alpha E)/(\beta+\gamma E)$.  This is just
an empirical fitting function, with no theoretical basis, but it gives a fairly
good fit to the data in Fig.~\ref{pvse}(a).  We can then differentiate this
function and calculate the observable polarization $P_{\text{obs}}(E)$ that
corresponds to the simulation results.

\begin{figure}
(a)\includegraphics[clip,width=3.2in]{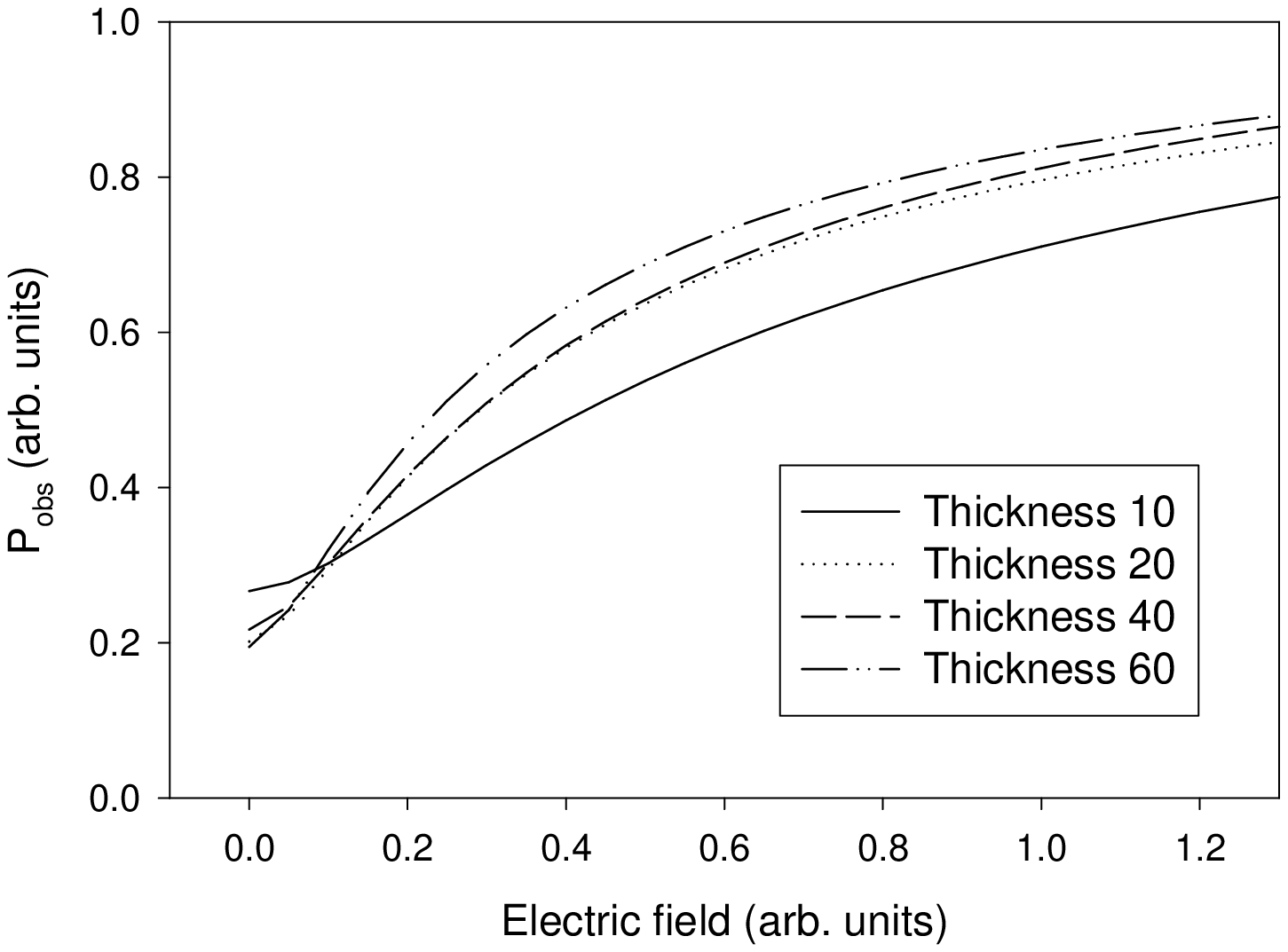}%

(b)\includegraphics[clip,width=3.2in]{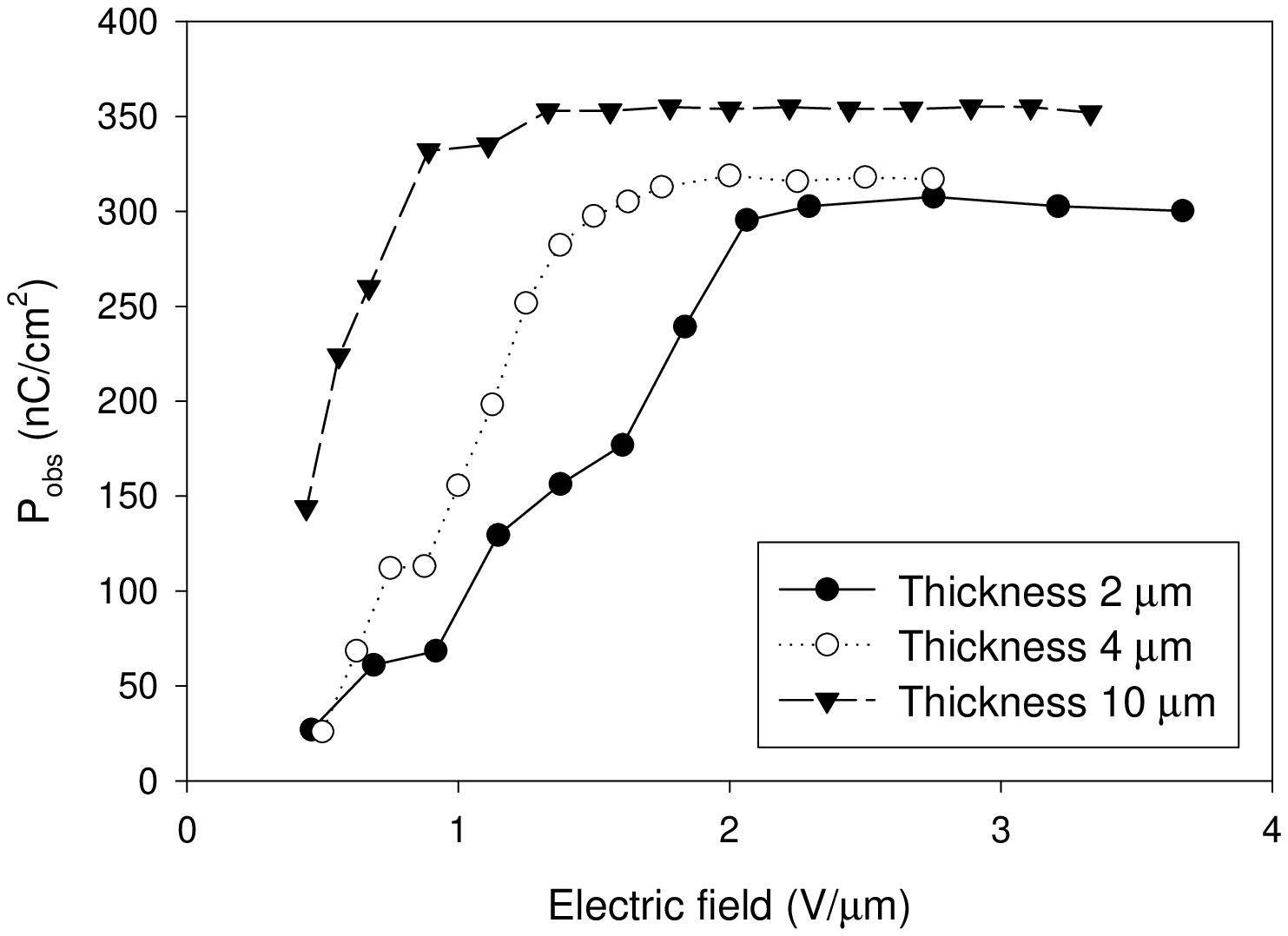}%
\caption{(a)~Simulation results for the observable polarization
$P_{\text{obs}}=P-E(dP/dE)$ in cells of four thicknesses.  (b)~Experimental data
for $P_{\text{obs}}$ in three cells~\protect{\cite{crandallprivate}}.}
\label{pobsvse}
\end{figure}

Figure~\ref{pobsvse}(a) shows the results for $P_{\text{obs}}(E)$ extracted from
the simulations at each system size.  This function is small for low $E$, then
increases towards its saturation level $P_{\text{obs}}=b$ at high $E$.  As
discussed above, thinner cells have a higher value of $P_{\text{obs}}$ at low
field, and thicker cells have a higher value of $P_{\text{obs}}$ at high field.
By comparison, Fig.~\ref{pobsvse}(b) shows sample experimental measurements of
$P_{\text{obs}}$ as a function of applied electric field in three cells with
asymmetric boundary conditions~\cite{crandallprivate}.  The material is 10PPBN4
(described in Ref.~\cite{naciri95}), and the temperature is 7 $^\circ$C below
the SmA-SmC transition.  Note that the experimental data show the same general
features as the theoretical plots.  In particular, we see the same crossover
between higher $P_{\text{obs}}$ in thinner cells at low field and higher
$P_{\text{obs}}$ in thicker cells at high field.  Thus, the simulation results
are consistent with the trends seen in experiments.

In conclusion, we have developed an approach for simulating the helix
winding/unwinding transition in SmC* liquid crystals.  This approach is based on
a minimal model for the free energy, which includes a chiral term that favors a
helical modulation of the director from layer to layer.  This bulk free energy
competes with surface effects and electric-field effects, which both favor a
uniform alignment of the director.  In zero field, the competition between the
bulk helix and the surface alignment leads to helix unwinding at a critical
thickness approximately equal to the helical pitch.  When an electric field is
applied, the field-induced alignment adds to the surface effects, and induces
helix unwinding even for thicker cells.  The electrostatic polarization is an
appropriate order parameter to quantify this field-induced unwinding, and our
simulation results for the polarization are consistent with experimental
measurements.

\begin{acknowledgments}
We would like to thank R. B. Meyer for helpful discussions and K. A. Crandall
for sharing his unpublished data with us.  This work was supported by the U.~S.\
Navy through Contract No.\ N00173-99-1-G015, and by the National Science
Foundation through Grant No.\ DMR-9702234 and DMR-0116090.
\end{acknowledgments}

\end{document}